\documentclass[prl,twocolumn,preprintnumbers,amsmath,amssymb]{revtex4}

\usepackage{graphicx}
\usepackage{dcolumn}
\usepackage{bm}

\begin{document}

\title{
Redshift of excitons in carbon nanotubes 
caused by the environment polarizability}
\author{Michael Rohlfing}
\email{Michael.Rohlfing@uos.de}
\affiliation{
  Fachbereich Physik, Universit\"at Osnabr\"uck, 
       Germany\\
}

\date{\today}

\begin{abstract}

Optical excitations of molecular systems can be modified
by their physical environment.
We analyze the underlying mechanisms within many-body perturbation 
theory, which is particularly suited to study non-local polarizability
effects on the electronic structure.
Here we focus on the example of a semiconducting carbon nanotube,
which observes redshifts of its excitons when the tube is touched
by another nanotube or other physisorbates.
We show that the redshifts
mostly result from the polarizability of the attached ad-system.
Electronic coupling may enhance the redshifts, but
depends very sensitively on the structural details of the contact.

\end{abstract}

\pacs{73.22.-f,78.67.Ch,71.15.Qe}



\maketitle

With improving precision of nanostructured sample preparation, 
controlled manipulation of optical excitations on the atomic scale
becomes possible, opening a path to nanometer photonics in
anorganic, organic, and pure-carbon semiconductor systems.
One prototypical material is given by
carbon nanotubes (CNT) and their excitons
\cite{Bac02,Bac03,Spataru04,Chang04,Maultzsch05,Finnie05,Wang06,%
Avouris07,Bondarev2009,Spataru10}.
Here the influence of the surrounding physical environment
is of high importance for the stability and control of excited states.
In this Letter we discuss (by many-body perturbation theory, 
MBPT \cite{Onida02,Rohlfing00}) two 
mechanisms by which excitonic states in CNT and in other nano-scale 
semiconducting systems can be manipulated by surrounding material.

In order to demonstrate the effects, we consider systems that
have recently been investigated experimentally.
The optical excitations of a CNT exhibit redshifts to lower
energy when the CNT is put in nitrogen \cite{Finnie05}
or when it is touched by another CNT \cite{Wang06}.
To our knowledge, a comprehensive understanding and 
analysis of these effects is still missing, 
hindering their systematic and widespread investigation.
We study these issues within 
MBPT \cite{Onida02,Rohlfing00},
which has become the standard approach to CNT 
excitons \cite{Spataru04,Chang04,Maultzsch05,Spataru10}.
We include the physical environment (here: another CNT or nitrogen), 
thus going far beyond those 
previous MBPT studies in which the CNT were considered in vacuum.
We believe that our approach and conclusions apply to many other 
materials, as well.

The spectrum of one system (1) might be shifted in energy 
by various mechanisms when contacted with another system (2).
We focus on two mechanisms which could be most relevant 
for chemically inert semiconducting systems:
(i) modification of the exciton energy in system 1 by 
electronic polarization of system 2, and
(ii) electronic coupling between systems 1 and 2.
We show below that the second effect (ii) depends heavily on the 
details of the contact geometry, thus requiring sample perfection
that may be too difficult to achieve for CNT.
The redshifts from the polarizability effect (i), on the other hand,
are much more robust and may fully explain the experimental findings 
\cite{Finnie05,Wang06}.
Throughout the paper we focus on {\em electronic} polarizability, i.e. 
nuclear motion or phonons and corresponding polarization are not 
considered.
Also, spectral shifts from changes of the intrinsic structure of the 
CNT (e.g., deformation or chemical modification)
are not considered in this paper, assuming that physisorption of
inert material leaves the CNT structure unaffected.

Both the fundamental gap and the excitonic 
binding energy are reduced when system 1 is contacted with other 
polarizable material 
\cite{Rohlfing03,Neaton06,Thygesen09,Thygesen11,Greuling11}.
However,  a homogeneous change
of the dielectric background would change both quantities by the same
amount (to lowest order), leaving the optical excitations unchanged.
%
Instead,
the shifting of optical excitations requires 
additional polarizability 
{\em with spatial inhomogeneity in closest vicinity of system 1}
(e.g. substrate surfaces, neighboring molecules, etc.).
In some cases, such additional polarizability might be described as
a continuous medium with a (flat or curved) surface, allowing to 
express its effects on system 1 by analytical expressions (e.g., 
solvent models or image-potential effects) \cite{Thygesen11}.
However, it is not clear if such an approach applies to the
highly anisotropic and inhomogeneous polarizability of an adjacent CNT.
Instead we include the additional polarizability with atomistic 
resolution in our MBPT approach \cite{Rohlfing10}.

\begin{figure}
\scalebox{0.60}{\includegraphics{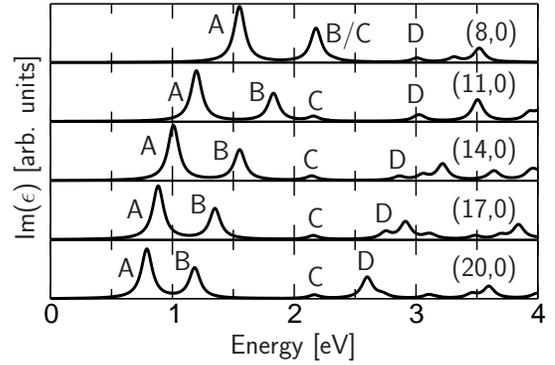}}
\caption{
\label{fig1}
Optical spectra of isolated ($N$,0) CNT
(for the electric field of the light along the tube).
Characteristic peaks are denoted as A, B, C, and D 
(cf. Fig. \ref{fig2} d,e).}
\end{figure}
Two technical specifications should be mentioned:
(i) We mostly employ a simplified "LDA+$GdW$"
version of MBPT \cite{Rohlfing10,Greuling11}.
While being somewhat less accurate (on an absolute energy scale) 
than a full GW/BSE calculation with 
RPA dielectric screening, LDA+$GdW$ still fully incorporates
all relevant aspects of the screening (atomistic resolution, 
local-field effects, and non-locality).
Our reference calculations within the conventional $GW$/BSE/RPA 
approach confirms the applicability of LDA+$GdW$ (see below).
(ii) 
Both in the GWA and the BSE part of LDA+$GdW$, we employ the same 
{\bf k}-point sampling (16 {\bf k}$_i$-points
from the first Brillouin zone, including the $\Gamma$ point)
for representing the screened Coulomb interaction $W$.
Since each {\bf k}$_i$ represents 
a sub-volume $V_i$ of reciprocal space,
we replace $W({\bf k}_i)$ by its average in $V_i$, i.e. by
$1/V_i \int_{V_i} W({\bf k}) d^3k$.
This guarantees the aforementioned cancelation of gap reduction and
electron-hole interaction reduction
when the dielectric background is changed or when polarizability
is added at large distance from the CNT.

%
Optical spectra for individual ($N$,0) CNT ($N$=8-20) are shown in 
Fig. \ref{fig1}.
For the (8,0) CNT, the first four optically active excitations (A-D) 
are found at 1.55 eV, 2.18 eV, 2.33 eV, and at 3.01 eV.
For the larger tubes,
peaks A and B move to lower excitation energy while
C and D remain in the visible range.
These LDA+$GdW$ results (for the (8,0) CNT) differ slightly from our 
full GW/BSE/RPA reference calculation,
which yields 1.60 eV, 2.05 eV, 2.42 eV and 3.16 eV for
peaks A-D.
A previous GW/BSE/RPA calculation \cite{Spataru04} 
yielded 1.55 eV and 1.80 eV
in comparison with experimental data of 1.60 eV and 1.88 eV 
\cite{Bac02,Bac03}.
The slight deviations of our LDA+$GdW$ data result from the 
approximations involved and from the employment of a model 
screening.
The screening \cite{Rohlfing10}
is based on the static RPA polarizabilites of the carbon atoms
(neglecting local-field effects) which amount to 
7-7.8 \AA$^3$ (depending on $N$) in the direction along the tube axis,
5.5-6.7 \AA$^3$ along the tube surface but perpendicular to the axis,
and 1.1-1.2 \AA$^3$ in radial direction.
Starting from these values, $\epsilon_{{\bf G},{\bf G}'}({\bf q},\omega)$
fully incorporates 
inhomogeneity, non-locality and local-field effects \cite{Rohlfing10}.


Starting from the spectra of Fig. \ref{fig1}, we now include in the
screening the polarizability of system 2.
\begin{figure}
\scalebox{0.60}{\includegraphics{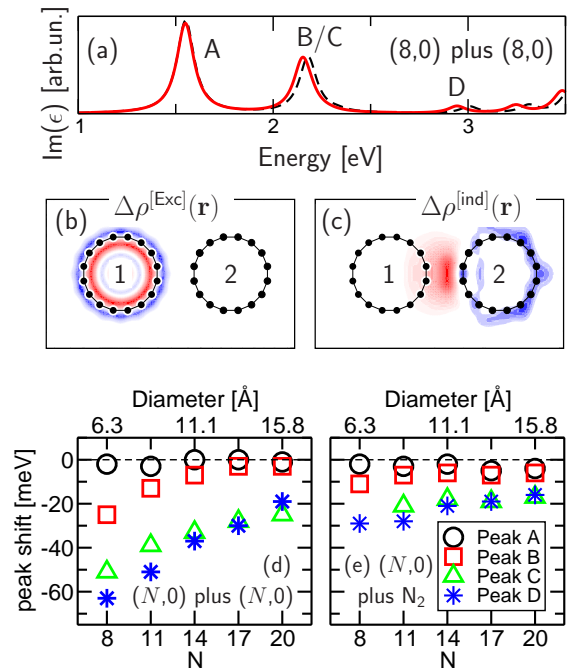}}
\caption{
\label{fig2}
Effect of {\em additional polarizability} on the optical spectrum of a
single CNT, {\em disregarding} electronic coupling.
(a) Spectrum of a (8,0) CNT while in contact
with another (8,0) CNT.
The isolated (8,0) CNT spectrum (Fig. \ref{fig1}a)
is included as dashed curve.
(b) Charge distribution
[$\Delta\rho^{[Exc]}({\bf r}) := \rho_v({\bf r})-\rho_c({\bf r})$]
of exciton D from panel a (blue: negative charge, red: positive charge).
(c) Induced charge distribution on the other CNT
(blue: negative charge, red: positive charge).
(d) Redshift of spectral peaks A-D of a ($N$,0) CNT when in
contact with another ($N$,0) CNT (i.e., two equal CNT).
(e) Redshift of spectral peaks A-D of a ($N$,0) CNT when being
covered by N$_2$ molecules [State C cannot be identified for $N$=8].
}
\end{figure}
Fig. \ref{fig2} a shows the effect on a (8,0) CNT
when another (8,0) tube is attached to it (at a
distance of 3.15 \AA).
All peaks are redshifted to lower excitation 
energy.
Note that the redshifts are significantly smaller than
the reduction of the fundamental gap and of the exciton binding energy
(both $\sim$0.3 eV).
Both effects largely cancel each other, yielding 
only a small net effect.
Our full GW/BSE/RPA reference calculation yields the same redshifts
to within 10 meV.

The redshift results from the polarization of CNT 2 when an
exciton on CNT 1 is excited \cite{Thygesen11}.
As illustration, Fig. \ref{fig2} b shows the change of electronic
charge 
[$\Delta\rho^{[Exc]}({\bf r}) := \rho_v({\bf r})-\rho_c({\bf r})$]
when exciton D is excited on CNT 1.
Since the conduction ($c$) states are closer to the vacuum level than 
the valence ($v$) states, the former extend farther into the vacuum, 
causing $\Delta\rho({\bf r})$ to be slightly positive inside CNT 1 and 
slightly negative outside.
This slight inhomogeneous charge distribution of the exciton leads
to a polarization of the material nearby (here: CNT 2), as shown
in Fig. \ref{fig2} c [induced charge density 
$\Delta\rho^{\sf [ind]}({\bf r})$].
The interaction between $\Delta\rho^{[Exc]}({\bf r})$
and $\Delta\rho^{\sf [ind]}({\bf r})$ finally redshifts the excitation.

Note that such effects are particularly important if
$\Delta\rho^{[Exc]}({\bf r})$ is non-zero at such positions {\bf r}
where system 2 has high charge susceptibility (caused by its own
electronic structure) and inhomogeneity.
This is mostly the case at distances of about 1-3 \AA\ from the
nuclei of system 2.
Here system 2 can be polarized by $\Delta\rho^{[Exc]}({\bf r})$ even 
if it carries no dipole.
For any exciton, $\Delta\rho^{[Exc]}({\bf r})$ must be non-zero 
somewhere (if not simply for the above-mentioned 
argument that electrons extend farther into vacuum than holes).
The effect described here should thus be of widespread relevance.

Since the range of the susceptibility of system 2 amounts to a few \AA,
only, the redshifts are quite sensitive to the distance between systems
1 and 2. 
We find that they decrease by 50 \% when the distance between
CNT 1 and CNT 2 increases by 1 \AA.
Significant redshifts thus require physisorption distance.
This would be different for excitons with charge-transfer dipole, for
which the polarizability effects are long-ranged and are generally 
stronger \cite{Thygesen11}.

From Figs. \ref{fig2} b and c it is clear that the polarizability
only affects a few atoms of the CNT, i.e. in the immediate
contact between system 1 and 2.
Since the excitons are equally distributed around the tube,
the amplitude of $\Delta\rho^{[Exc]}({\bf r})$ at each atom scales
like $1/N$.
An exciton of a larger-diameter CNT thus causes weaker polarization 
of CNT 2 and observes smaller redshifts, as can be seen in our
data for all five ($N$,0) CNT (each contacted with another ($N$,0) CNT)
in Fig. \ref{fig2} d.
Wang {\em et al.} measured redshifts of 30-50 meV \cite{Wang06} 
(in the visible spectrum) for CNT with diameter of 18-19 \AA\, (maybe 
about 15 \AA\, in our notation of nucleus-to-nucleus distances).
Our results at that diameter are comparable to these data (although 
slightly smaller).

Effects independent of CNT diameter can be expected when the entire 
surface is covered by polarizable adsorbates.
In the experiments by Finnie {\em et al.} \cite{Finnie05}
physisorbed N$_2$ was the most likely explanation for the observed
redshift of 20-30 meV \cite{Finnie05}.
We find that N$_2$ physisorbs on graphene (flat-lying at a distance 
of $\sim$3 \AA, slightly depending on the adsorption site)
with a DFT-LDA binding energy of $\sim$80 meV (in DFT-LDA).
We have included N$_2$ molecules in the CNT calculations.
The physisorption on the CNT was modeled by a Lennard-Jones potential
based on our N$_2$-graphene DFT-LDA results, complemented by an 
N$_2$-N$_2$ intermolecular potential \cite{Galassi94}.
Room-temperature molecular dynamics (MD) then yields realistic 
structures
with the entire CNT surface covered by N$_2$.
Subsequent LDA-$GdW$ calculations (averaged over 10 MD snap shots)
yield the redshifts shown in Fig. \ref{fig2} e.
The physisorption of N$_2$ does in fact yield redshifts that 
depend much less on tube diameter.
However, the polarizability of N$_2$ is much smaller than
that of an adjacent CNT, leading to significantly weaker
redshift effects.

\begin{figure}
\scalebox{0.60}{\includegraphics{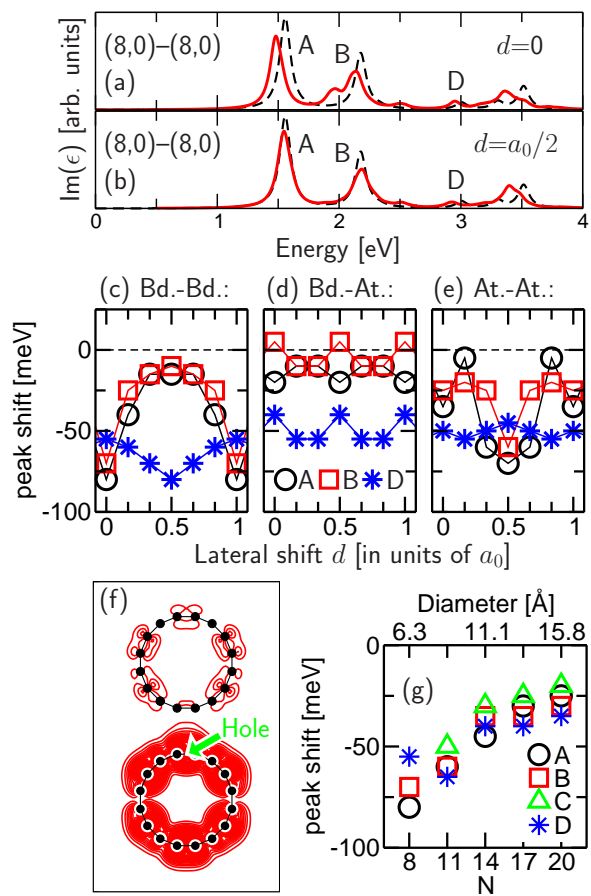}}
\caption{
\label{fig3}
Effect of {\em electronic coupling} between two CNT on their 
spectra.
(a,b) Spectrum of two coupled (8,0) CNT for
two contact structures (Bond-Bond configuration with $d$=0 and 
$d$=$a_0/2$; cf. panel c).
(c-e) Redshift of the spectral peaks A,B, and D of two
coupled (8,0) CNT, depending on the contact structure.
[Peak C cannot be identified in many of the spectra for $N$=8.]
(f) Electron distribution (relative to the hole) for exciton A of 
panel a.
(g) Redshift of A-D of a ($N$,0) CNT when in
contact with another ($N$,0) CNT (all in Bond-Bond configuration 
with $d$=0, cf. panel a).
}
\end{figure}

So far, the electronic degrees of freedom of the adjacent
object were not considered.
This is certainly valid for the physisorption of N$_2$, which has
HOMO (LUMO) states far below (above) those of the CNT.
For the coupling between two CNT, on the other hand, significant
inter-tube interaction among electrons and holes is expected,
calling for calculations in which CNT 2 is fully included
(not only its polarizability).
It turns out that (i) for some of the excitons the effects are much 
more drastic than the effects from the polarizability alone, and (ii) 
the coupling depends very sensitively on the contact geometry.
Fig. \ref{fig3} a+b shows spectra for
a pair of (8,0) CNT in two different configurations.
In both structures, bonds of the two CNT face each other (cf.
Fig. \ref{fig2} b+c and Fig. \ref{fig3} f), but for panel b 
tube 2 was laterally shifted by half a lattice constant 
($d$ = $a_0/2$ =2.13 \AA), leading to completely different redshifts.
The dependence of the redshifts on the contact geometry
is summarized in Fig. \ref{fig3} c-e.
In addition to the lateral shift $d$, CNT 2 was rotated by 
11.25$^\circ$ for panel d, such that bonds of CNT 1 face atoms
on CNT 2 (''Bond-Atom'').
For panel e, both tubes were rotated by 11.25$^\circ$ such that atoms 
face atoms along the connection line between the tubes (''Atom-Atom'').
The resulting redshifts of peaks A and B range from --80 meV to
zero. 
They are extremely sensitive to the slightest rotation or 
shift of the tubes, which strongly change 
the coupling-matrix elements between the atomic wave-function
components (that are highly sensitive to the orientation of the
atoms to each other).
Peak D, on the other hand, always shows a redshift of about
--60 meV already observed in Fig. \ref{fig2}, i.e. no significant
shift beyond the polarizability effect (which is, of course, 
automatically included).
Such high sensitivity is {\em not at all} found for the polarizability 
effect (Fig. \ref{fig2}), which is the same (to within 1 meV) for all 
rotational angles and shifts.
Apparently, the polarizability effect is much more robust than
electronic coupling.

This high sensitivity of the electronic coupling raises
the question if this effect would be relevant for experimental 
observations, like Ref. \cite{Wang06}.
Exactly the same contact geometry would be
required over many lattice constants, which seem unlikely.
Even the basic requirement of having two CNT with identical chirality
(and thus lattice constant) may not be fulfilled.
Slightly different lattice constants of the two CNT and consequently
non-commensurable contact geometry would immediately reduce the
electronic coupling effect.
We therefore believe that mainly the polarizability effects as discussed
before can be held responsible for the measured redshifts.
However, a slight increase (on average) of the redshifts 
might result from the electronic coupling.

Nonetheless a closer analysis of the electronic 
coupling is helpful and may guide further studies in which
the required perfect contact quality can be achieved. 
In particular, the mere size of redshifts of up to 80 meV (in addition
to the polarizability effect) is interesting.
At first glance the interaction between two
identical excitons $|\Psi_1\rangle$ and $|\Psi_2\rangle$ on identical 
CNT 1 and 2 should mainly lead to superpositions 
$|\Psi_1\rangle\pm|\Psi_2\rangle$ with only slight changes of the 
excitation energy; furthermore, the energy shifts should be more or less
symmetric to the red and the blue (however, one of the coupled excitons
might get the full dipole strength and the other one become dark).
In fact, our analysis of test states like 
$|\Psi_1\rangle\pm|\Psi_2\rangle$
do show very small spectral shifts for peaks A and B.
The significant redshift of A and B (for some geometries) 
results from a totally different effect, i.e. a spatial spill-out 
of the excitons across the CNT contact.
Schematically, $|\Psi_1\rangle \hat{=} |v1$$\to$$c1\rangle$ is composed 
from holes ($v$) and electrons ($c$) on CNT 1
($|\Psi_2\rangle \hat{=} |v2$$\to$$c2\rangle$ on CNT 2, respectively).
Electronic coupling adds charge-transfer contributions
$|v1$$\to$$c2\rangle$ and $|v2$$\to$$c1\rangle$ (which are not contained
in states like 
$|\Psi_1\rangle\pm|\Psi_2\rangle$), thus forming an 
exciplex state (similar to excimer states among atoms or molecules).

The exciplex nature can be visualized by the electron-hole
correlation function, e.g.  Fig. \ref{fig3} f.
For the hole on CNT 1, the electron is mostly found on the same tube
but also has significant amplitude (about 10\%) on CNT 2 (from
the $|v1$$\to$$c2\rangle$ contribution).
The admixture of such charge-transfer contribution constitutes 
a reduced quantum confinement (QC) between electron and hole across the
tubes, compared to the QC on a single tube.
QC tends to increase excitation energies (e.g., in semiconductors);
here we find the reversed effect of reduced excitation energy
(i.e., redshift) due to reduced QC.
[In excimers this effect is responsible for significant
interatomic binding.]
Similar to the size dependence of the polarizability effect 
(Fig. \ref{fig2} d), this exciplex redshift becomes smaller for
the larger CNT, as shown in Fig. \ref{fig3} g (all for Bond-Bond 
contacts with $d$=0).


In conclusion, we have shown that electronic polarizability of
neighboring systems can redshift exciton states of carbon nanotubes.
Here the exciton's charge-density distribution induces charge density
in the neighboring system.
This mechanism is particularly effective when the excited system 
is very close to the neighboring system, e.g. at physisorption 
distance.
This should be relevant not only for carbon nanotubes (which were
taken as an example in the present study), but also for other 
molecules, polymers, etc. in contact with chemically inert systems
that exhibit electronic polarizability.
In addition to the polarizability effect, electronic coupling between
the systems can significantly enhance the redshifts.
However, very precise control of the contact structure would be
required for electronic coupling, since it depends very sensitively
on the atom positions of the two components relative to each other.
If the contact is disturbed by periodicity mismatch, thermal 
fluctiation, or other uncontrolled features, electronic coupling
might be immediately destroyed.
The polarizability effect, on the other hand, is very robust with
respect to contact details.
Here we investigated both mechanisms within many-body perturbation
theory, which allows to address both the non-local
polarizability effect and electronic coupling.

We thank the John von Neumann Institut f\"ur Computing at 
FZ J\"ulich for computational resources.

\end{document}